\documentclass[referee]{aa}

\usepackage{graphicx}
\usepackage{txfonts}
\usepackage{lipsum}
\usepackage{subcaption}         
\usepackage{lscape}             
\usepackage{placeins}           
\usepackage{txfonts}
\usepackage{gensymb}
\usepackage[version=4]{mhchem}
\usepackage{hyperref} 
\usepackage{amsmath}
\usepackage{amssymb}
\usepackage{natbib}
\usepackage{xcolor}
\usepackage{threeparttable}
\usepackage{bbding}
\usepackage[english]{babel}
\usepackage{epstopdf}

\begin{document} 
   \title{Synthesis of organo-phosphorous species in space: the reaction of P$^+$ with \ce{C2H2}}


   \author{Matteo Michielan\inst{1}\email{matteo.michielan@unitn.it}
          \and
          Jorge Alonso de la Fuente\inst{2}\email{jorge.alonso@iff.csic.es} 
          \and
          Roland Thissen\inst{3,4}\email{roland.thissen@universite-paris-saclay.fr}
          \and
          Christian Alcaraz\inst{3,4}\email{christian.alcaraz@universite-paris-saclay.fr}
          \and
          Nicolas Solem\inst{3}\email{nicolas.solem@universite-paris-saclay.fr}
          \and
          Cristina Sanz Sanz\inst{5}\email{cristina.sanz@uam.es}
          \and
          Susana G\'omez Carrasco\inst{6}\email{susana.gomez@usal.es}
          \and
          Miroslav Pol\'a\v{s}ek\inst{7}\email{miroslav.polasek@jh-inst.cas.cz}
          \and
          Alexandre Zanchet\inst{2}\fnmsep\corrauth{alexandre.zanchet@csic.es}
          \and
          Marcelino Ag\'undez \inst{2}\email{marcelino.agundez@csic.es}
          \and
          Daniela Ascenzi\inst{1}\fnmsep\corrauth{daniela.ascenzi@unitn.it}
          }

\institute{Dipartimento di Fisica, Università di Trento, Via Sommarive 14, 38123 Trento, Italy
        \and 
        Instituto de  F\'{\i}sica  Fundamental, CSIC, Serrano 123, 28006, Madrid, Spain
        \and
        Universit\'e Paris-Saclay, CNRS, Institut de Chimie Physique, UMR 8000, 91405 Orsay, France
        \and
        Synchrotron SOLEIL, L’Orme des Merisiers, 91192 Saint Aubin,Gif-sur-Yvette, France
        \and
        Departamento de Qu\'{\i}mica F\'{\i}sica Aplicada (Unidad Asociada de I+D+i al CSIC), Modulo 14, Universidad Autonoma de Madrid, Madrid 28049, Spain
        \and
        Departamento de Qu\'{\i}mica F\'{\i}sica, Facultad de Ciencias Qu\'{\i}micas, Universidad de Salamanca, 37008 Salamanca, Spain
        \and
        Heyrovsk\'y Institute of the Czech Academy of Sciences,
        Prague, Czechia
        }

   \date{Received XX XX, 2026; accepted XX XX, 2026}

   \abstract
   {Despite its low cosmic abundance, phosphorus is a bioessential element whose prebiotic availability and incorporation into biomolecules remain open questions. Only few molecules containing phosphorous-carbon bonds have so far been detected in astronomical environments (CP, CCP, HCP). Although more complex species, analogous variants of N-containing organics have been proposed, their formation and destruction pathways remain poorly constrained due to incomplete astrochemical networks. Gas-phase ion–molecule chemistry involving \ce{P+} may play an important role in the formation of organo-phosphorous compounds, particularly in protostellar shocks and outflows. Moreover, reactions of ground-state \ce{P+} with closed-shell neutrals provide valuable benchmarks for studying spin-forbidden reaction mechanisms.}
   {The reaction between \ce{P+} and acetylene (\ce{C2H2}) is investigated through a joint experimental and theoretical study, with the aim of revising the rate coefficients currently reported in astrochemical databases and resolving an existing discrepancy, whereby calculated rate coefficients exhibit a strong temperature dependence but are more than one order of magnitude lower than the experimentally measured rate at 300 K.
   }
   {The reaction was investigated experimentally by measuring absolute cross sections (CSs) and branching ratios (BRs) as a function of collision energy. Experiments were complemented  by a theoretical investigation combining high-level electronic structure calculations with a refined capture model that properly accounts for the probability of intermediate-complex formation by including, in addition to charge-induced dipole, higher order components. Channel-specific rate coefficients as a function of temperature, over the range 10-5000 K, were derived.}
   {The reaction \ce{P+} + \ce{C2H2}, at astrochemically relevant temperatures, leads mostly to \ce{HCCP+} plus H (BR=99\%), with \ce{CCP+} plus \ce{H2} being a minor channel (BR=1\%), occurring due to intersystem crossing (ISC) from the triplet to the singlet potential energy surfaces. The obtained total reaction rate coefficient at 300 K ($1.1\times10^{-9}$ cm$^3$s$^{-1}$) is in good agreement with SIFT measurement ($1.3\times10^{-9}$ cm$^3$s$^{-1}$), but differ from a pure Langevin trend by showing an increase with decreasing temperature ($1.3\times10^{-9}$ and $8.9\times10^{-10}$ cm$^3$s$^{-1}$ at 10 and 5000 K, respectively).
   }
   {The reaction of \ce{P+} with acetylene should be considered in astrochemical models in which phosphorus can be released in the gas phase as a cation, e.g. from the energetic processing of icy interstellar grains due to shocks. The reaction leads to linear \ce{HCCP+} as the main reaction product and relevant interstellar isomer, from which the observed species (\ce{CP}, \ce{CCP}) could be obtained via dissociative recombination with electrons.}


\titlerunning{P+ with \ce{C2H2}}
\authorrunning{}

\maketitle
\nolinenumbers

\section{Introduction}
In recent years, the detection and the chemistry of phosphorus-bearing (hereafter P-bearing) species in space have drawn remarkable attention from the astrochemical and astrobiological communities \citep{Fontani_2024_rev}. This growing interest arises mainly from the well known fundamental role of phosphorus in the origins of life research.
Phosphorus, along with sulfur (S), is one of the third-row elements that, despite their low abundances in the Universe, plays a crucial role as a biogenic element. In fact, P-bearing compounds such as phosphates (\ce{PO_4^{3-}}), are essential in the formation of complex and structurally stable biomolecules such as deoxyribonucleic acid (DNA), ribonucleic acid (RNA), the main carriers of genetic information, adenosine triphosphate (ATP), and phospholipids. 
In a wider perspective, the detection of P-bearing molecules in several environments of the ISM and planetary or exoplanetary atmospheres offers valuable insights into prebiotic chemistry, as these molecules may have been delivered to early Earth from extraterrestrial sources via asteroids and comets. The recent discovery of phosphorus-rich grains, with no clear resemblance to other meteorite minerals, during the sample return mission from asteroid Ryugu provides a further piece in this direction, suggesting that such phosphorus delivery may have critically contributed to the reaction pathways of organic matter towards biochemical evolution \citep{Pilorget_2024}.
The prebiotic origin of \ce{P} is an open issue in origin of life research: a key factor governing the formation of organophosphorus species on Earth and exoplanets is the abundance and availability of \ce{P} in a form suitable for undergoing phosphorylation reactions necessary to incorporate phosphate groups into biomolecules \citep{pasek_thermodynamics_2020, walton_phosphorus_2023, walton_chemical_2026}. In the early Earth, the limited availability of water-soluble phosphate minerals, coupled with their low reactivity toward the formation of phosphoester (P-O-C) bonds, has led to the proposal that other species containing P in a reduced oxidation stage may have been present \citep{pasek_evidence_2013, hao_active_2025}.
Such species (e.g., schreibersite, an iron-nickel phosphide (Fe, Ni)$_3$P, exhibits higher reactivity and solubility in water, making them more plausible sources of \ce{P} in living organisms \citep{pantaleone2024atomistic, pantaleone2025prebiotic, gull_phosphorylation_2025}. Additionally, reduced \ce{P} species could react with \ce{NH3} leading to the nitrogenous versions of phosphates, amidophosphates \ce{O=P-NH2} group), that have been shown to facilitate phosphorylation reactions \citep{gibard_geochemical_2019, gull_prebiotic_2023}.

In the present-day Solar photosphere, the relative abundance of P with respect to H is $\sim 3 \times 10^{-7}$ \citep{Asplund_2009}, that is, about 2 orders of magnitude lower than that of sulfur (S), 
and this explains why, while more than 40 S-bearing molecules have been identified in the ISM (https://cdms.astro. uni-koeln.de/classic/molecules), only a small number of P-bearing molecules have been clearly identified so far, mostly in star-forming regions and in the envelopes of evolved stars, but also in other environments, such as extragalactic sources and Solar system objects.

Unlike the numerous and ubiquitous N-containing interstellar organic species, when N is replaced by isoelectronic phosphorus, the number of organo-phosphrous species so far observed is limited to CP, CCP, and HCP, all detected in the circumstellar envelopes of carbon-rich evolved stars \citep{guelin_1990,Milam_HCP_CP_2008, Halfen_CCP_2008, Agundez_HCP_2007}, in addition to a tentative detection of cyanophosphaethyne (NCCP) \citep{agundez_new_2014}. Other more complex species with phosphorus–carbon multiple bonds, highly unstable under normal laboratory conditions on Earth, could potentially form and stabilize in rarefied astrophysical environments. These include phosphaalkynes and phosphaalkenes such as \ce{HC3P}, \ce{CH2CHCP} and \ce{CH2PH}, the P analogues of cyanoacetylene,  vinyl cyanide and methanimine \citep{lawzer_2025}; \ce{C2H5CP}, the P equivalent of ethyl cyanide, whose gas phase rotational spectrum was recently characterized in the laboratory \citep{bonah_2024}; \ce{c-C5H4P}, the P variant of pyridine, and larger polycyclic aromatic phosphorus heterocycles \citep{fioroni_2019,oliveira_2021}. Although these species have yet to be detected by radioastronomical observations, the rapid progresses in high-resolution laboratory spectroscopy and increase telescope sensitivity are paving the way for their future astronomical detection, making reliable chemical data increasingly important for their identification and inclusion in astrochemical models.

The detection of only a limited number of P-bearing species in the ISM makes it challenging to constrain both the gas-phase elemental abundance of phosphorus and its dominant chemical pathways \citep{Recio:26}. While various chemical networks for P-containing molecules have been developed, including a sequence of gas phase ion-molecule and neutral–neutral reactions
\citep{thorne_1984, Adams_1990, Millar_1991, Charnley_1994, mackay2001phosphorus, Rivilla_2016, Jimenez-Serra_2018, Chantzos_2020, delaConcepcion_2021, Sil_2021}, only a small number of the proposed reaction rates or cross sections have been studied experimentally and in some cases the reaction pathways remain poorly constrained or unexplored.

A prominent role may be played by the gas phase reprocessing in protostellar shocks and outflows, where atomic \ce{P} can be present also as \ce{P+}, given the lower ionization energy of \ce{P} (10.36 eV) compared to that of its second row counterpart, N (14.5 eV).  In fact atomic phosphorus is detected in the gas-phase in diffuse clouds in the form of \ce{P+} \citep{Savage_Pabund_1996, Jenkins_P_abund_1986, Jura_P_obs_1978}, and \ce{P+} is also detected in the diffuse ISM of external and more distant galaxies \citep{Lebouteiller_P+_2013, Lebouteiller_P+_2006, Friedman_P+_2000}. 
Ion chemistry of \ce{P+} could then be relevant to explain the efficient synthesis of molecules containing P-O, P-N, and P-C bonds, highlighting the relevance of the results presented in this work.

The \ce{P+} + \ce{C2H2} reaction has been previously studied using the Selected Ion Flow Tube (SIFT) technique \citep{Smith_1989, Adams_1990}: a total rate coefficient of $1.3\times 10^{-9}$ cm$^3$ s$^{-1}$ at 300 K is obtained, with 95$\%$ of the reactive flux producing \ce{HCCP+} + \ce{H} and the remaining 5$\%$ producing the adduct \ce{PC2H2+}. These values are included as such in the KInetic Database for Astrochemistry (KIDA) \citep{KIDA2015, KIDA_2024} with a rate coefficient independent from temperature, assuming a Langevin capture model due to the absence of dipole moment for \ce{C2H2}. In the UMIST database for Astrochemistry \citep{UMIST2022} only the \ce{HCCP+} + \ce{H} channel is reported with a rate coefficient equal to $1.24\times 10^{-9}$ cm$^3$ s$^{-1}$ and no temperature dependence following the same assumption.
As already observed for previous systems \citep{MichielanP+_water, michielan2025experimental} Langevin capture models may be an inadequate assumption to describe reaction rates from experiments performed only at room temperature. Furthermore, experiments such as SIFT are performed at relatively high pressure and are likely to overestimate the production of adducts by stabilizing reaction intermediates by collisions with the buffer gas. 

From a theory point of view, the reaction profile of \ce{P+}+\ce{C2H2} reaction has been first established by \citet{largo_theoretical_1995}, in which a detailed study of the reaction intermediates was performed at MP2/6-31G* level of theory. They reached the conclusion that \ce{HCCP+} + H is the only exothermic product assuming a reaction on the lowest triplet potential energy surface (PES).
In a more recent work, \citet{cimas_computational_2012} reported energy profiles on the singlet and triplet PES at the CCSD(T)/aug-cc-pVTZ//MP2/cc-pVTZ level of theory, and they showed that the CCP$^+$+H$_2$ product was  also exothermic when considered in its singlet state, and may be reached through a spin-forbidden mechanism. In this work, a calculated rate coefficient as a function of temperature using RRKM theory was also provided. Unlike Langevin model assumption, the calculated rates exhibit a strong temperature dependence, but the estimation of the rate at 300 K is found about one order of magnitude lower than the one measured in the SIFT experiment, casting doubts on both experimental and theoretical results. 

To provide new clues, we have therefore decided to reassess the title reaction by performing a new experimental determination of the total reactive cross sections (CSs) and branching ratios (BRs) as a function of the collision energy in the 0.1 – 10 eV range using the guided ion beam (GIB) technique. The experiment was performed at low pressure in the cell to guarantee near single-collision regime. In addition, new theoretical calculations are performed considering excited electronic states to complete the picture provided by the previous theoretical studies. A refined capture model, improved from the standard Langevin one is also employed to properly account for the formation of the first complex intermediate, a barrierless process that can not be described properly by the RRKM theory. From the experimental CSs, and with the insights provided by theory, the rate coefficients as a function of temperature in the 10-5000 K range are inferred.

\section{Experimental Methodology}
\label{sec:exp_met}
Data on the reaction \ce{P+} + \ce{C2H2} have been collected using the Guided Ion Beam (GIB) setup CERISES \citep{cunha_de_miranda_reactions_2015, Alcaraz:04} coupled to the VUV DESIRS beamline \citep{Nahon:12} at the French synchrotron radiation facility SOLEIL. The set-up is a GIB tandem mass spectrometer composed of two octopoles (O) and two quadrupole mass filters (Q) in a quadrupole-octopole-octopole-quadrupole (QOOQ) configuration, that allows for the investigation of bimolecular reactions of mass-selected ions. The set-up and production of the parent ion beam have been already detailed in \cite{de_la_fuente_michielan_2026}, while additional technical details can be found in \cite{Zanchet:24, Zanchet:25} and \cite{Richardson:24}, so only a brief summary is given here.

Phosphorus cations (\ce{P+}) are produced via dissociative photoionization of the \ce{PCl3} precursor at $8.3\times 10^{-6}$ mbar in the ion source at selected photon energies ($E_\mathrm{phot}$). Upon mass selection with Q1, the centre-of-mass (CM) collision energy ($E_\mathrm{CM}$) at which P$^+$ collides with the \ce{C2H2} target gas -introduced in the scattering cell (O1) at a pressure of $3.1\times 10^{-5}$ mbar - is controlled by the potential difference between the ion source and scattering cell and the retarding field potential method is used to determine the zero of the kinetic energy scale \citep{teloy} . 
Product ions (as well as parent ions) are collected and guided (O2) into the second quadrupole (Q2) for mass-selection prior to detection by an electron multiplier.
Absolute reaction CSs are then extracted from the measured ion yields and the absolute pressure of the target gas, which is set so that the single-collision regime is ensured.
The effective cell length is determined through the calibration reaction Ar$^+$ + D$_2$ $\rightarrow$ ArD$^+$ + D \citep{Ervin_1985}. 

\ce{P^{+}} has two low-lying metastable excited states, \ce{^{1}D_2} and \ce{^{1}S_0}, located 1.10 and 2.67 eV above the ground state, respectively \citep{NIST_ASD}. Both states, if generated by dissociative photoionization, will have sufficiently long radiative lifetimes (in the range of tens of seconds for \ce{^{1}D} \citep{Huang_1985, Czyzak_1963} and in the range of $\sim$0.4 s for \ce{^{1}S} \citep{nahar_2025, Czyzak_1963} to reach the scattering cell and contribute to the reactivity with \ce{C2H2}.
Tuning the photon energy for dissociative photoionization of \ce{PCl3} and performing additional ion-molecule experiments is pivotal to assess the best experimental conditions to have a \ce{P+} beam containing exclusively the ground electronic state \ce{^{3}P}. Additional experiments on the reactivity of \ce{P+} with \ce{N2} to quantify the absence of metastable excited states of \ce{P+} is presented in Appendix \ref{sec:appendix_A}

The lowest appearance energy (AE) for \ce{P+}(\ce{^{3}P}) from \ce{PCl3} is observed at 15.5$\pm0.1$ eV photon energy, corresponding to the ion-pair dissociation channel giving \ce{P+}(\ce{^{3}P}) + \ce{Cl_2^-} + Cl, while a second ionization threshold, with a much higher ionization efficiency, is detected and attributed to a "complete" dissociation giving \ce{P+}(\ce{^{3}P}) + 3\ce{Cl} with a calculated dissociation threshold of 20.5 eV \citep{de_la_fuente_michielan_2026}. At higher photon energies the generation of excited metastable states is thermodynamically feasible: the thresholds for production of \ce{P+}(\ce{^{1}D}) are calculated at 16.6 eV and 21.6 eV for the ion-pair and "complete" dissociation channels, respectively. The corresponding thresholds to yield \ce{P+}(\ce{^{1}S}) are at 18.2 eV and 23.2 eV.
However, this does not imply that dissociative ionization of \ce{PCl3} will surely yield metastable states of \ce{P+}, since the photofragmentation process depends on the shape of the molecular cation PESs accessible upon photon absorption and connectivity to the dissociation channels (transition states may be present).    

\section{Theoretical Methodology}
\subsection{Electronic structure calculations}

Electronic structure calculations were performed to explore the interactions between \ce{P+} cation and \ce{C2H2} in the entrance channel and to shed light on the open channels from the different possible products of reaction in several electronic states. 
For the entrance channel, the ground state of phosphorus cation is \ce{^{3}P}, which present three different electronic states that are strictly degenerate when spin-orbit interaction is not considered. Since, in the experiment, \ce{P+} is produced by dissociative photoionisation of \ce{PCl3}, we can assume that all the fine levels of \ce{P+}(\ce{^{3}P}) are equally populated. It is therefore necessary to explore not only the interaction potential of the ground triplet state, but also from the two first excited triplet states. In addition, since spin-orbit effects are expected to be non negligible for third row atoms, it is also interesting to explore the behaviour of the singlet states. The entrance channel was modelled  considering the equilibrium geometry of \ce{C2H2}, which was kept frozen in a linear geometry allowing the use of the C$_s$ symmetry group. The interaction potential of acetylene with several electronic states of \ce{P+}, the three first triplets (1$^3A'$, 1$^3A''$ and 2$^3A''$) and the two first singlets (1$^1A'$ and 1$^1A''$), have been computed at CASPT2 level \citep{Werner:96}. The active space employed consists in 14 electrons in 9 orbitals, while the 7 orbitals associated to inner shells of P and C atoms were kept frozen. In addition, the long range potential was also calculated at CCSD(T) level \citep{Deegan-etal:94} for the 1$^3A''$ state.

To determine the reaction enthalpies, reactants and product channels in their ground state were optimized at CCSD(T) level. Due to the large vibrational zero point energy (ZPE) of acetylene, MP2 harmonic frequencies were also considered to include the ZPE correction.
To determine the energies of the products in their excited electronic states, geometry optimization of the different fragments (isolated) were performed for several isomers of each fragment for different singlet and triplet electronic states. The CASPT2 relative energies of the fragments in their excited electronic state were then summed to the CCSD(T) energies of the respective ground states to estimate their reaction enthalpies. 
All the calculations were performed using the AVTZ basis set in \citet{Dunning:89} within the MOLPRO package \citep{MOLPRO}.

\subsection{Capture model}

In order to go beyond the Langevin model, which consider a long range potential based exclusively on a charge-induced dipole interaction, we employ here a one-dimensional capture model similar to the one employed in the work of \citet{delMazo:24}. The concept is similar to the Langevin model, in the sense that it is applicable for reactions with no barrier in the entrance channel with a very large reaction probability if the intermediate complex forms. The assumption made in our one-dimensional model is that reactants will always reorient towards the most favourable orientation and will thus follow the most attractive potential energy curve (PEC). In the case of barrierless reactions, the only way to impede the reaction is the centrifugal barrier arising from the orbital angular momentum of the two colliding fragments. To consider this effect, it is common to use an effective potential which is function of the distance $R$ between the two fragments. In the quantum mechanics formalism, the effective potential is  defined as: 

\begin{equation}
V_{eff}(R)=V(R)+J(J+1)/2\mu R^2
    \label{eq:effpot}
\end{equation}
\noindent
where $V(R)$ is the interaction potential, $J$ is the total angular momentum and $\mu$ is the reduced mass of the fragments. For a simple implementation, all the terms have to be considered in atomic units (distances in Bohr, energy in hartree and mass referred to the mass of the electron). 
Applying Eq.\ref{eq:effpot} for different values of $J$, it appears that the position and height of the effective barrier maximum changes. Following the capture model philosophy, if we consider that all the collisions with translational energies larger than the effective barrier will lead to reaction, then it is possible to relate the position of the effective barrier maximum to the maximum impact parameter, $b_{max}$, leading to reaction for energies corresponding to the effective barrier height. By knowing the maximum impact parameter for each energy, we can then apply the classical mechanics definition of the cross section and simply calculate the CS as $\sigma=\pi b_{max}^2$. 

In this work, we considered the CCSD(T) potential energy curve which provided accurate description of the long range interaction as $V_{LR}(R)$. This curve was then fitted to a multipolar expansion of the form $V_{LR}(R)= -c_3/R^3-c_4/R^4-c_5/R^5-c_6/R^6$ to consider properly the charge-quadrupole, charge-induced dipole, quadrupole-induced dipole and induced dipole-induced dipole interaction. It was found necessary to employ all these terms to reproduce properly the CCSD(T) interaction potential. We should note here that the widely used Langevin model only considers the $-c_4/R^4$ term, as it allows an easy and direct estimation  of the CS. The model employed here goes beyond that, and provide the theoretical higher limit of the CS as function of collision energy.

\section{Results}
\subsection{Experimental cross sections (CSs) and branching ratios (BRs)}

\begin{figure}[h]
   \centering
   \includegraphics[width=\hsize]{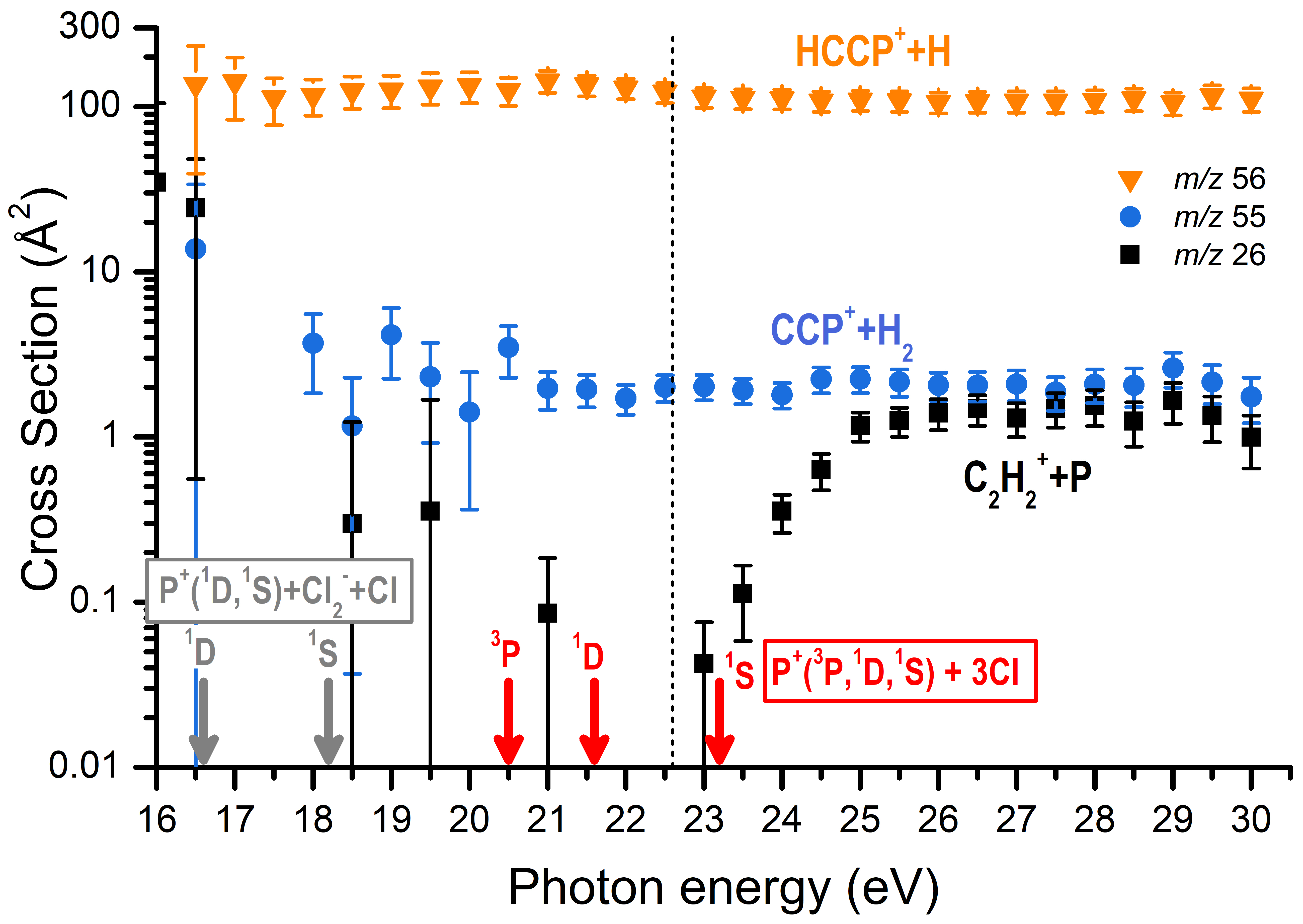}
   \caption{Absolute CSs for the reaction of \ce{P+} with \ce{C2H2}, measured at $E_\mathrm{CM}$ = 0.14 eV as a function of the photon energy. Products are detected at \textit{m/z} 56 (\ce{HCCP+}, orange triangles) and \textit{m/z} 55 (\ce{CCP+}, blue circles). From $E_\mathrm{phot}\approx$23 eV an additional product is observed at \textit{m/z} 26 (\ce{C2H2+}, black squares). The vertical dashed black line indicates $E_\mathrm{phot}$ value at which CS trends as a function of $E_\mathrm{CM}$ have been measured. Grey and red arrows indicate the calculated thermodynamic thresholds for production of \ce{P+} in ground and excited states, as detailed in the text.}
         \label{fig:P_C2H2_hv_0.14eV}
   \end{figure}

\begin{figure}[h]
   \centering
   \includegraphics[width=\hsize]{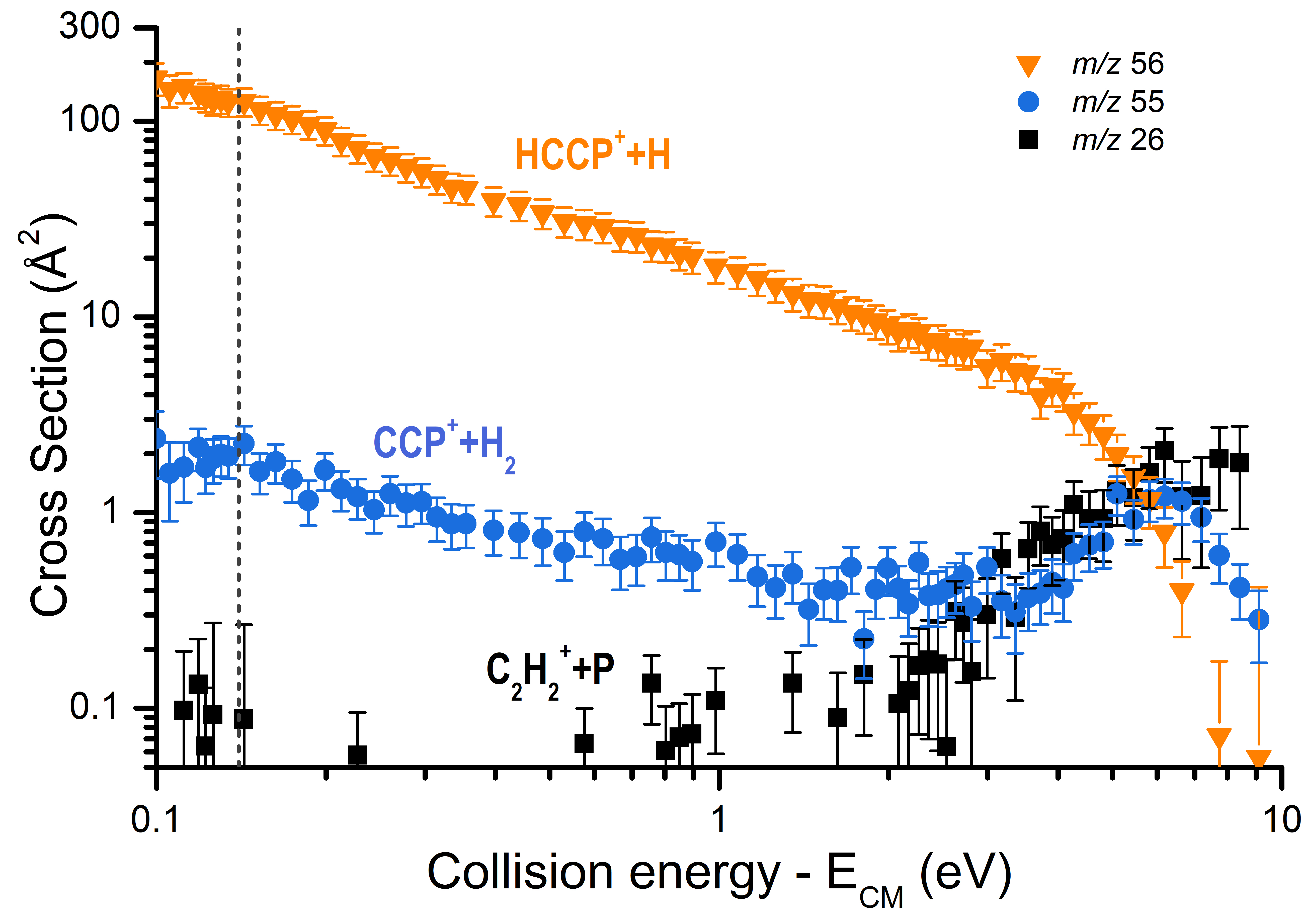}
   \caption{Absolute experimental CSs as a function of collision energy, at $E_\mathrm{phot}$ = 22.6 eV. Products are detected at \textit{m/z} 56 (\ce{HCCP+}, orange triangles) and \textit{m/z} 55 (\ce{CCP+}, blue circles). From $E_\mathrm{CM} \approx$ 1-2 eV the additional charge transfer product is observed at \textit{m/z} 26 (\ce{C2H2+}, black squares). The vertical dashed line indicates $E_\mathrm{CM}$ value at which CS trends as a function of $E_\mathrm{phot}$ have been measured.}
         \label{fig:P_C2H2_ecm_22.6eV}
\end{figure}

Absolute CSs as a function of photon energy are obtained from the intensity of the product and parent ions, measured at a fixed collision energy ($E_\mathrm{CM}$ = 0.14 eV), and are shown in Fig.\ref{fig:P_C2H2_hv_0.14eV}, while CSs measured as a function of $E_\mathrm{CM}$ at $E_\mathrm{phot}$ = 22.6 eV are reported in Fig.\ref{fig:P_C2H2_ecm_22.6eV}. 

Two product ions are observed at \textit{m/z} 56 and 55 in the whole photon and collision energy ranges explorable with our technique, and identified with reaction channels  (\ref{eq:second_P_C2H2}) and (\ref{eq:first_P_C2H2}).  An additional channel at \textit{m/z} 26 and corresponding to the charge transfer is also observed (\ref{eq:third_P_C2H2}), but with an appearance energy thresholds in terms of photon or collision energies. 
\begin{align}
\ce{P+ + C2H2} &\rightarrow \ce{HCCP+ + H} &  & \textit{m/z}\hspace{0.2cm} 56 & \label{eq:second_P_C2H2} \\
               &\rightarrow \ce{CCP+ + H2} &  & \textit{m/z}\hspace{0.2cm} 55 & \label{eq:first_P_C2H2} \\
               &\rightarrow \ce{C2H2+ + P} &  & \textit{m/z}\hspace{0.2cm} 26 & \label{eq:third_P_C2H2}
\end{align}

Reaction enthalpies, which have been estimated from {\it ab initio} calculations
and are reported in Table \ref{tab:enthalpy_P_C2H2} together with data available in the literature \citep{ATcT, NISTchem, cimas_computational_2012}, are compatible with such observations.
Although some small differences, attributable to different methodologies, are observed between our calculated enthalpies and those from previous works, all calculations agree on the fact that channels (\ref{eq:second_P_C2H2}) and   (\ref{eq:first_P_C2H2}) are exothermic and may be formed even at low collision energies, while the charge transfer channel lies higher in energy.

\begin{table*}
\centering
\caption{Reaction enthalpies ($\mathbf{\Delta}H^0$) for the products of the reaction of \ce{P^+(^3P)} with \ce{C2H2}}
\label{tab:enthalpy_P_C2H2}
\begin{threeparttable}
\begin{tabular}{l c c c c}
\hline\hline
\noalign{\smallskip}
\textbf{Reaction Products} & \textit{m/z}\tnote{a} & $\Delta H^0$, eV\tnote{b} & PES \tnote{c}  & $\Delta H^0$, eV (literature) \\
\noalign{\smallskip}
\hline\hline
\ce{HCCP+}($X^2\Pi$) + \ce{H}($^2$S) & 56 & $-0.307$   & 1$^3A'$, 1$^3A''$, 1$^1A'$, 1$^1A''$ & $-0.069$ \tnote{d}  \\
\ce{CCP+}($X^1\Sigma$) + \ce{H2}($^1\Sigma^+_g$) & 55  & $-0.209$ & 1$^1A'$ & $-0.182$ \tnote{d}\\
\ce{c-HCCP+}($X^2A'$) + \ce{H}($^2$S) & 56 & $+0.0145$   & 1$^3A'$, 1$^1A'$ & $+0.082$ \tnote{d}  \\
\ce{CPC+}(X$^1B_2$) + \ce{H2}($^1\Sigma^+_g$) & 55 &  $+0.084$ & 1$^1A'$ &   \\
\ce{CPC+}(a$^3B_2$) + \ce{H2}($^1\Sigma^+_g$) & 55 &  $+0.584$ & 1$^3A'$ & $+0.859$ \tnote{d} \\
\ce{CCP+}($a^3\Pi$) + \ce{H2}($^1\Sigma^+_g$) & 55 &  $+0.805$ &  1$^3A'$, 1$^3A''$ & \\
\ce{C2H2+}($X^2\Pi_u$) + \ce{P}($^4S$) & 26 & $+1.072$ &  1$^3A'$, 1$^3A''$ & $+0.91$ (VEE $+1.003$)\tnote{e} \\
\ce{CPC+}(b$^3B_1$) + \ce{H2}($^1\Sigma^+_g$) & 55 &  $+1.820$ & 1$^3A''$ &  \\
\ce{CCP+}($b^3\Pi$) + \ce{H2}($^1\Sigma^+_g$) & 55 &  $+2.227$ &  2$^3A'$, 2$^3A''$ & \\
\ce{CPC+}(c$^3B_1$) + \ce{H2}($^1\Sigma^+_g$) & 55 &  $+3.330$ & 2$^3A''$ &  \\
\ce{c-HCCP+}($A^2A''$) + \ce{H}($^2$S) & 56 & $+3.436$   & 2$^3A''$, 2$^1A''$ & $+3.803$ \tnote{f}  \\
\hline\hline
\end{tabular}
\begin{tablenotes}
\footnotesize
\item[a]: The $m/z$ value associated to the ionic fragment in each product channel
\item[b]: From {\it ab initio} calculations performed in this work
\item[c]: The PESs correlating to each product channel are shown
\item[d]: from \citet{cimas_computational_2012}
\item[e]: from NIST \cite{NISTchem}
\item[f]: from \citet{largo_theoretical_1995}
\end{tablenotes}
\end{threeparttable}
\end{table*}

CSs for \textit{m/z} 56 and 55 show a steady behaviour as a function of $E_\mathrm{phot}$ (see Fig.\ref{fig:P_C2H2_hv_0.14eV}), indicating the absence of effects due to the internal energy of the parent ion (i.e. its electronic state) . The most abundant channel, at \textit{m/z} 56, represents $\sim99$\% of the overall reactivity and is associated with the exothermic formation of the \ce{HCCP+} ion and \ce{H} as the neutral counterpart. The channel related to \textit{m/z} 55, \ce{CCP+} + \ce{H2}, makes up the remaining $\sim1$\%. Channel (\ref{eq:third_P_C2H2}) at \textit{m/z} 26 is due to charge exchange and it shows a clear threshold at $E_\mathrm{phot}\approx$ 23 eV. Considering the endothermicity of charge exchange from \ce{P+} in the ground \ce{^{3}P} state (see Table \ref{tab:enthalpy_P_C2H2}), the appearence of \ce{C2H2+} is ascribable to the production of a certain amount of metastable \ce{P+} in the \ce{^{1}S} state, in good agreement with the thermodynamic threshold, at 23.2 eV, for  dissociative ionization of \ce{PCl3} into \ce{P+}(\ce{^{1}S}) + 3Cl,  (see red arrow in Fig. \ref{fig:P_C2H2_hv_0.14eV}). 
Furthermore, the absence of charge transfer channel (\ref{eq:third_P_C2H2}) at 
$E_\mathrm{phot}$ below 23 eV, indicates that the production of \ce{P+} in the \ce{^{1}D} is either absent or present in such a low amount that is not sufficient to give a detectable yield of products.
We note that data points associated with charge transfer at energies lower than 23 eV are due to noise and are not statistically significant. In fact, at photon energies lower than $\approx$ 20.5-21 eV, the uncertainty of the measured CSs are large, and show a scattered behaviour rather than a trend, as a consequence of the low photoion yield for \ce{P+} from dissociative photoionization of \ce{PCl3}, as discussed in \citet{de_la_fuente_michielan_2026}.

A photon energy equal to 22.6 eV is chosen to measure absolute CS as a function of collision energy (Fig.\ref{fig:P_C2H2_ecm_22.6eV}), since it represents an optimal balance between sufficient reagent ion yields and a negligible presence of \ce{^{1}D} as a contaminant in the \ce{P+} beam.
As already evidenced from previous theoretical papers, the occurrence of channel (\ref{eq:first_P_C2H2}) leading to \ce{CCP+} is endothermic on the triplet PES.
A non zero CS for \ce{CCP+} at low collision energy (see Fig. \ref{fig:P_C2H2_ecm_22.6eV}) suggests either the presence of \ce{P+} in one of the excited singlet states - for which channel (\ref{eq:first_P_C2H2}) becomes the most exothermic channel (by about 1.37 eV from \ce{P+}(\ce{^{1}D})) - or a contribution from intersystem crossing (ISC) from the triplet to the singlet PES, as also proposed in \citet{cimas_computational_2012}.

Our data are crucial to disentangle among the two possibilities, by comparing the trends of \ce{CCP+} and \ce{C2H2+} product channels. Considering that the charge exchange channel is exothermic for the \ce{^{1}D} and \ce{^{1}S} states, if a contamination was present in the ion beam the \ce{C2H2+} product should be observed at low collision energies, but this is clearly not the case (Fig. \ref{fig:P_C2H2_ecm_22.6eV}). 
In the data as a function of photon energy (Fig. \ref{fig:P_C2H2_hv_0.14eV}), while the \ce{CCP+} products is present at all photon energies from 19 eV onward (as said at lower $E_\mathrm{phot}$ we can not measure low yield products due to the small amount of \ce{P+} generated), the \ce{C2H2+} appears only above about $\sim23$ eV, corresponding to the AE of \ce{P+}(\ce{^{1}S}) formed together with 3 Cl atoms. The presence of \ce{CCP+} product below such energy (e.g. at $E_\mathrm{phot}=22.6$ eV at which CS as a function of collision energy are measured), is a clear indication that \ce{P+} in the \ce{^{1}D} is, if not absent, present in a negligible amount in the ion beam, not sufficient to yield a detectable product signal. Therefore the production of \ce{CCP+} is expected to derive from ISC to the singlet PES, similarly to what observed for the \ce{PO+} product in the reaction of \ce{P+} with \ce{D2O} \citep{MichielanP+_water}.

The CSs associated with \ce{HCCP+} and \ce{CCP+}  products (Fig. \ref{fig:P_C2H2_ecm_22.6eV}) both show a decrease with increasing collision energy up to $\approx$ 2.5 eV, the expected behaviour from an exothermic and barrierless reaction. The charge transfer channel is absent at low collision energies and start to be appreciable only above 2 eV, compatible with the literature endothermicity from \ce{P+}($^3P$) and low efficiency close to threshold. The measured trend of this channel is a further confirmation of the absence of metastable \ce{P+}(\ce{^{1}D}), since if it were present, the \textit{m/z} 26 product would have shown a non zero CS at low collision energies due to the 0.19 eV exothermicity (0.91 -1.10 eV)
To confirm the absence of barrier, the potential energy curves of the entrance channel of the P$^+$+C$_2$H$_2$ reaction have been computed at CASPT2 level and are shown in the top panel of Fig. \ref{fig:PECs}. It appears clearly that for one of the three electronic states correlating to P$^+(^3P)$, a very stable complex can form without any barrier, consistent with the findings of \citet{cimas_computational_2012}.

\subsection{From cross sections to rate coefficients}

\begin{figure}[h]
    \centering
    \includegraphics[width=\hsize]{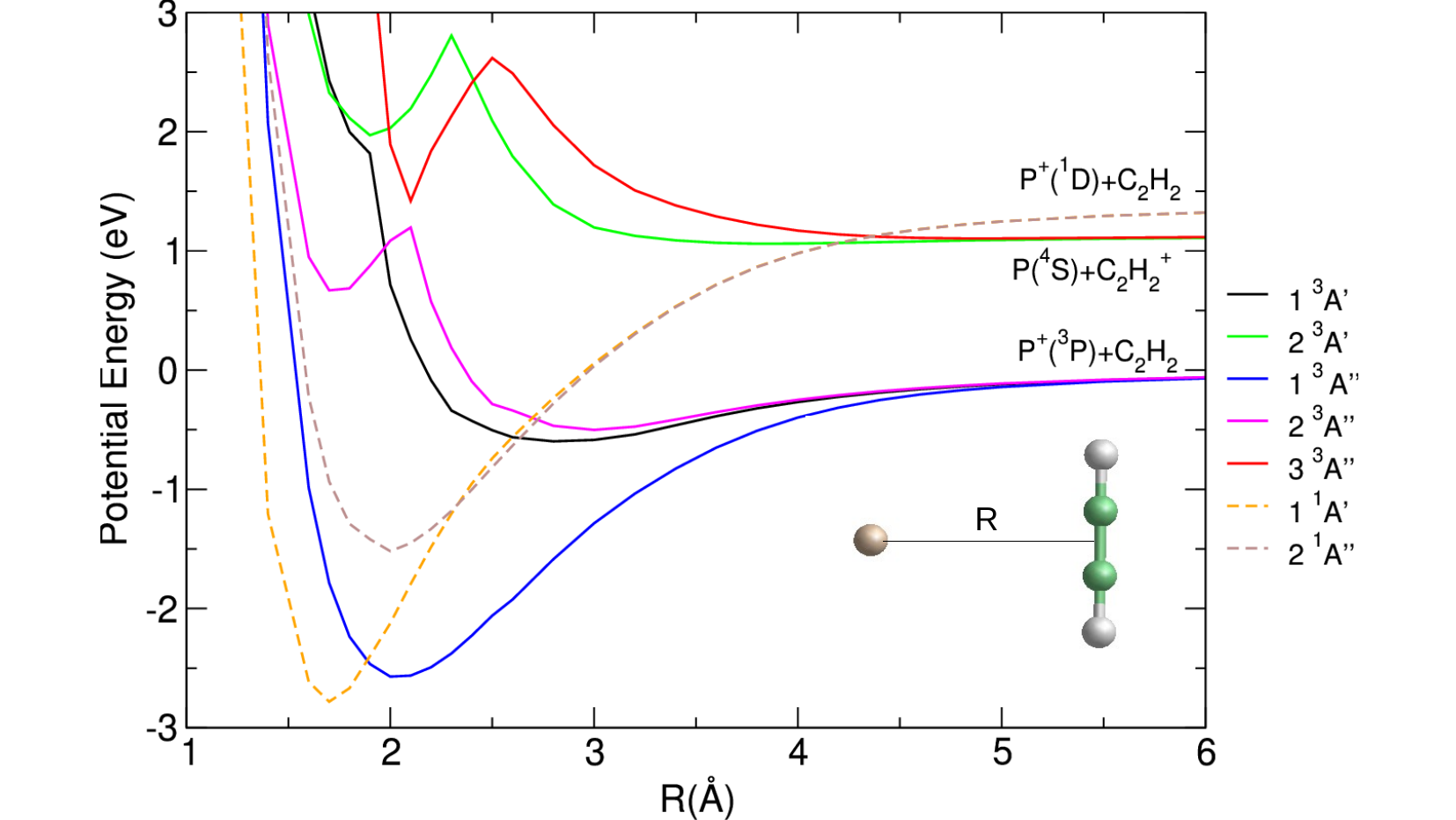}
    \vspace{0.1cm}
    \includegraphics[width=0.8\hsize]{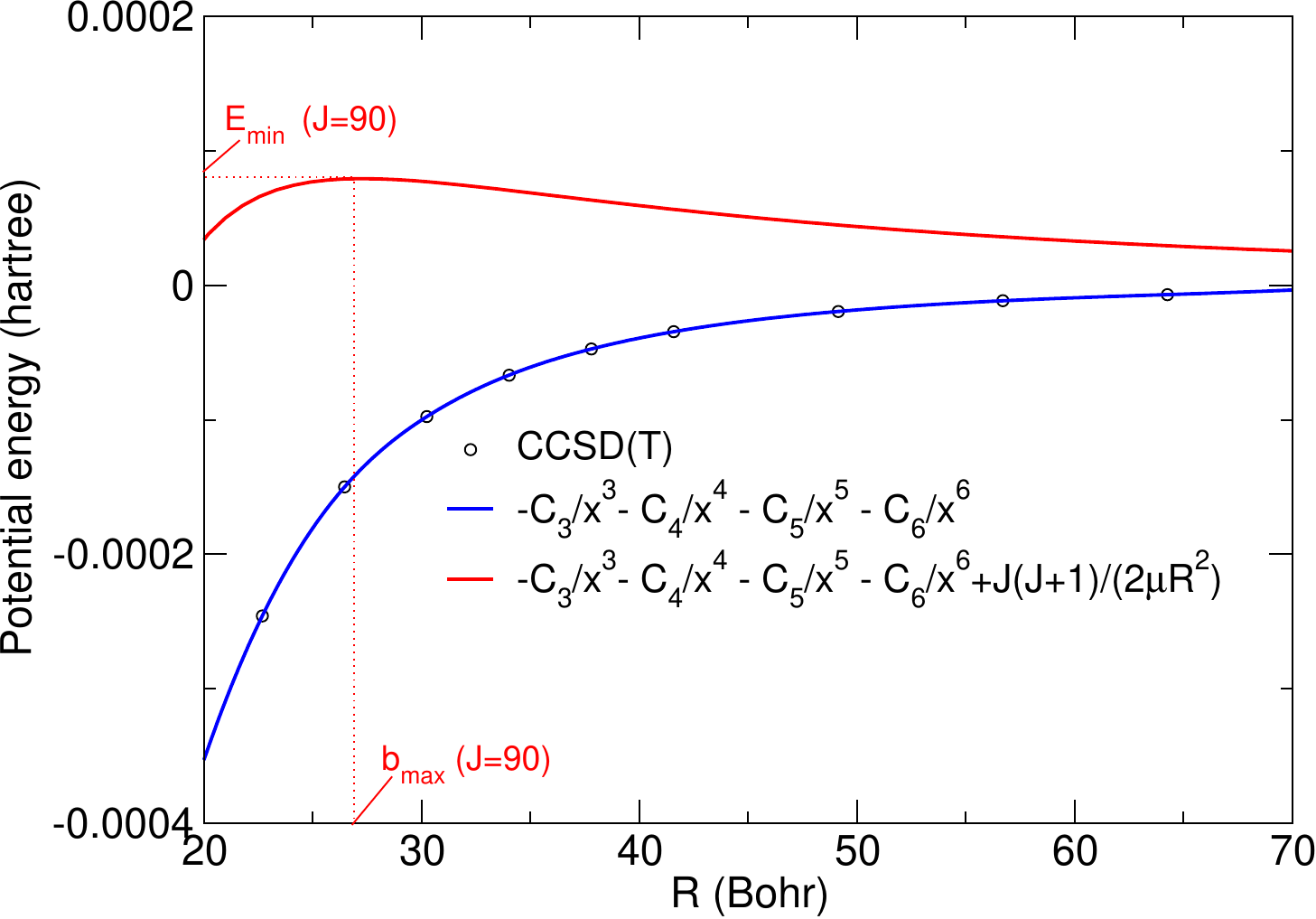}
    \caption{\textit{Top}: CASPT2 interaction potentials in eV for several electronic state of P$^+$+C$_2$H$_2$ for a T-shape approach (see inset). \textit{Bottom}: CCSD(T) long range potential in hartree. For illustration purpose, the effective potential considering a total angular momentum J=90 is also shown with its associated b$\_{max}$.}
    \label{fig:PECs}
    \end{figure}

\begin{figure}[h]
   \centering
   \includegraphics[width=\hsize]{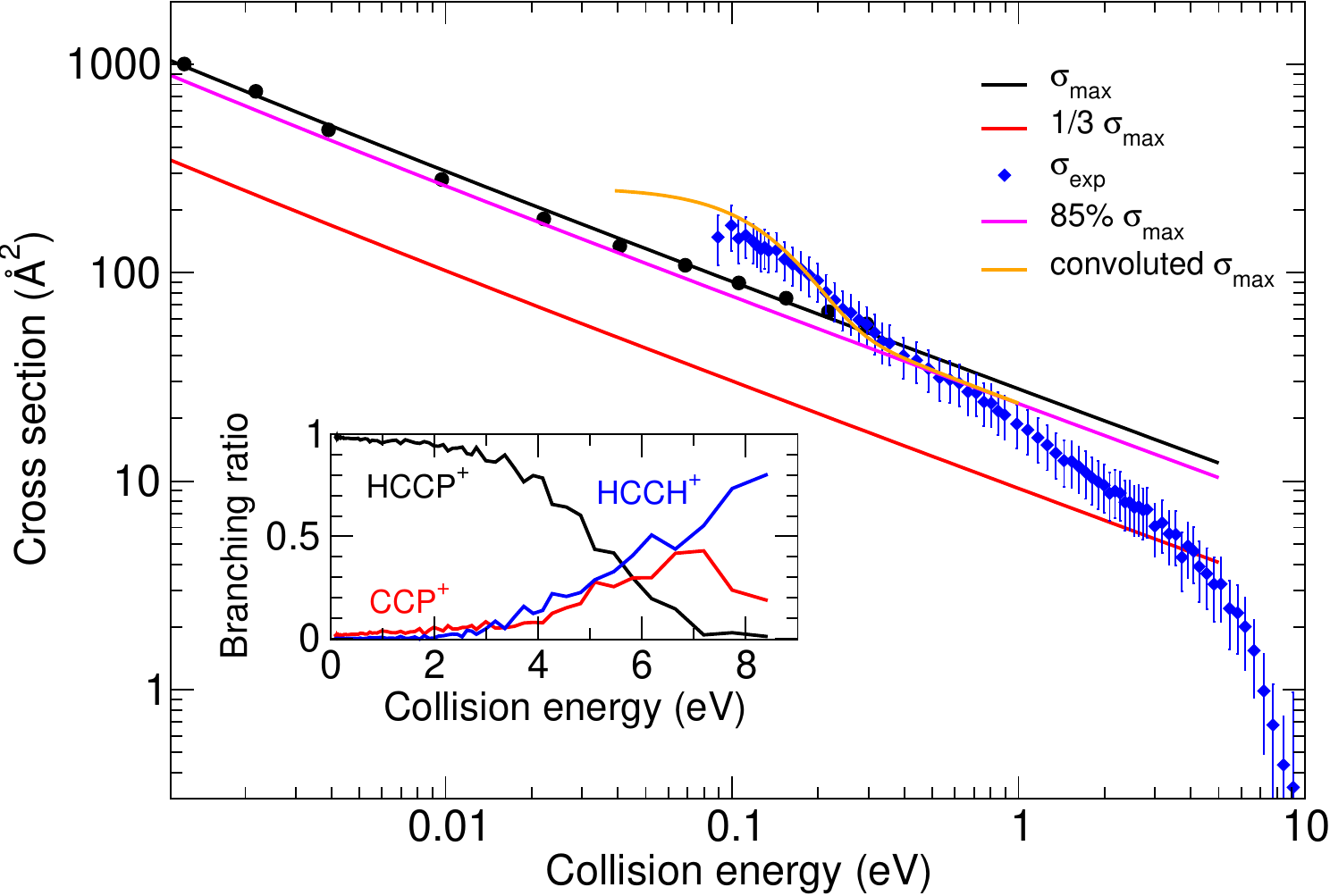}
   \caption{CSs for the reaction \ce{P+} with \ce{C2H2} as a function of $E_\mathrm{CM}$ at $E_\mathrm{phot}$ = 22.6 eV: the blue filled diamonds labelled $\sigma_{exp}$ represent the measured total CS, the black curve labelled  $\sigma_{max}$ represents the theoretical maximum value of the CS and is obtained fitting the black dots (see text), the red and magenta curves are fractions of $\sigma_{max}$, and the orange curve represent a convolution  of the CS (magenta curve) considering a dispersion of 0.1 eV. In addition, an inset representing the evolution the BR as a function of collision energy is also shown.}
   \label{fig:totCS}
   \end{figure}

As a result of the finite ion energy distribution of the reagent ion beam, the experimental CSs as a function of the ion energy differ from the true CS as a function of the relative collision energy.
The energy dispersion of the beam is estimated to be around 0.1 eV.
As a result, the measured cross sections below 0.2 eV are expected to be affected and present larger uncertainties. To recover the proper energy behaviour of the CS at low collision energy, we applied the improved capture model based on the CCSD(T) long range of the P$^+$--C$_2$H$_2$ interaction shown in the bottom panel of Fig. \ref{fig:PECs}.
The expected values of the maximum CS calculated for several values of $J$ are represented as black dots in Fig. \ref{fig:totCS}. The theoretical points can then be fitted to an analytical expression of the form:
\begin{equation}
    \sigma(E) = A\cdot {E^{-n}} + B \cdot {E^{-m}}
    \label{eq:sigmatrue}
\end{equation}

The best fit is found for $A=0.97$, $n= 0.971$, $B= 26.76$ and $m=0.5$, and it is shown as a black solid line in Fig. \ref{fig:totCS}. 
Interestingly, the second part of Eq. \ref{eq:sigmatrue} recovers the Langevin expression, however an additional term makes the CS increase faster when reaching the ultra low energies limit.
In Fig. \ref{fig:totCS}, when comparing the measured total CS $\sigma_{exp}$ (blue filled diamonds) to the estimated maximum limit  $\sigma_{max}$, it appears that  $\sigma_{exp}> \sigma_{max}$ below $\sim0.15$ eV, confirming an impact of the dispersion of the energy beam on the measurements at low collision energy.
At 0.2 eV,  where the energy spread is not expected to play a role, $\sigma_{exp}=0.85\times \sigma_{max}$ (represented by the magenta curve). Applying a convolution considering a dispersion of 0.1 eV \citep{Ervin_1985} to $0.85\times \sigma_{max}$, the convoluted CS (orange curve) reproduces quite well the experimental ones.
We will thus make the assumption that below 0.2 eV, the real CS will follow the trend of $0.85\times \sigma_{max}$, while above 0.2 eV, it will be equal to the measured CS $\sigma_{exp}$.

The CS derived this way is overall slightly steeper than the Langevin prediction, suggesting a contribution of the charge-quadrupole interaction particularly effective at low collision energies. 
Indeed, as  mentioned previously, the CS at low energies depends on the long range potential behaviour. 
In general, the charge-induced dipole interaction is considered as the dominant term for ion-non polar molecule interactions at large distances.
As such, in the standard Langevin model it is assumed that the long range interaction is given only by the charge-induced dipole interaction, which is proportional to $R^{-4}$.  
C$_2$H$_2$ present however a large quadrupolar moment and the long range behaviour of the charge-quadrupole interaction, which is  proportional to $R^{-3}$, is likely to dominate the long range interaction, allowing in addition some reorientation before the effective- ion-molecule collision. 
This non-standard Langevin behaviour of the CS was also observed in the case of the S$^+(^2D)$+H$_2$ reaction \citep{Zanchet:25}, also attributable to the quadrupolar moment of H$_2$.

It is also interesting to point out that the branching ratio (BR) of products changes significantly with collision energy, as shown in the inset of figure~\ref{fig:totCS}. At low collision energies, the main product is \ce{HCCP+} which accounts for 99\% of products while \ce{CCP+} counts for 1\%, ten times more than predicted by \citet{cimas_computational_2012}. The BR of \ce{HCCP+} slowly decreases with increasing collision energy and stops to become the principal product between 5 and 6 eV. 
Since the rate coefficient is calculated on an average of collision energies, providing the total rate with the BR of products is not convenient.
Instead, we prefer to fit independently the CS associated to each product with a proper extrapolation at low energy, and calculate the corresponding rate coefficients using the relation:
\begin{eqnarray} 
k(T)=  \left ( {8 k_BT \over \pi \mu} \right)^{1/2} (k_BT)^{-2} \int_0^\infty 
\sigma(E) E~ \exp \left({-{E \over k_B T}}\right) dE
\label{KdeT}
\end{eqnarray}
with $\mu$ the reduced mass of \ce{P+}+\ce{C2H2} and $k_B$ the Boltzmann constant.

\begin{figure}[h]
   \centering
   \includegraphics[width=\hsize]{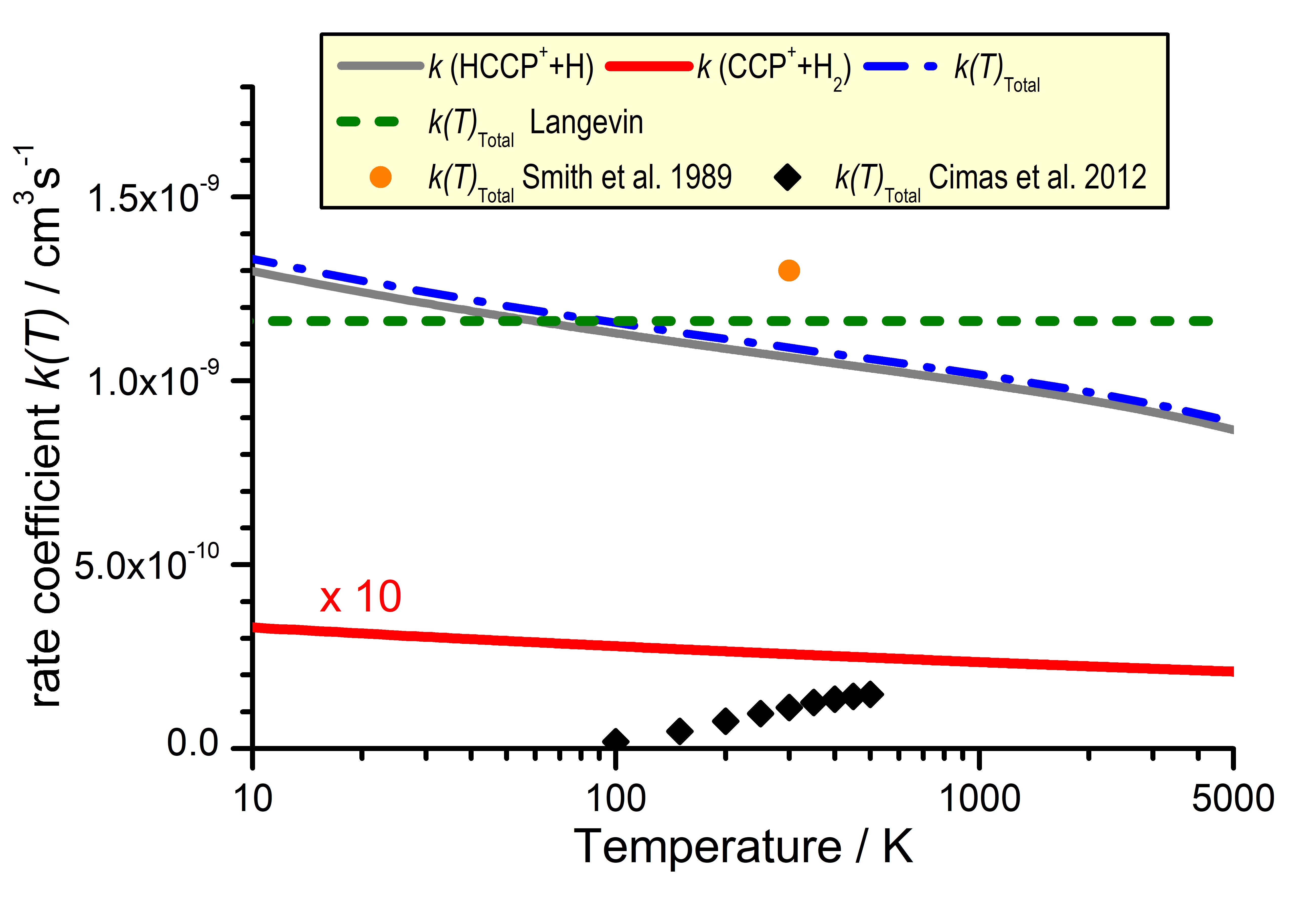}
   \caption{Partial and total rate coefficients $k(T)$ as a function of temperature (in the 10-5000 K range) for the \ce{P+} + \ce{C2H2} reaction as extrapolated from the experimental CS data plus improved capture model. The red and grey curves refer to channels (\ref{eq:second_P_C2H2}) and (\ref{eq:first_P_C2H2}) respectively, while the dashed-dotted blue line is the total $k(T)$, sum of partial rates (relative uncertainties, not shown for clarity reasons, are estimated as $\pm35$\%). The red curve is multiplied by a factor 10 to ease viewing. The dashed green curve is the Langevin value, the black diamonds are the total rates as calculated in \cite{cimas_computational_2012} and the orange dot is the total $k(T)$ from SIFT experiments \citep{Smith_1989, Adams_1990} (relative uncertainty, not shown, is $\pm30$\%).
      }
      \label{fig:ratesP+_C2H2}
   \end{figure}

From the fitted CS in a wide collision energy range and using Eq. \ref{KdeT}, the rate coefficients as a function of temperature $k(T)$ associated to \ce{HCCP+} and \ce{CCP+} products are reported in Fig. \ref{fig:ratesP+_C2H2}. 
In Table \ref{table:param} we report the parameters that should be used to incorporate the reaction rate coefficient for the two open channels into astrochemical models, following the modified Arrhenius equation usually employed in KIDA \citep{KIDA_2024} and defined as: 
\begin{equation}
k(T) = \alpha  \left(\frac{T}{300}\right) ^\beta e^{-\gamma/T} 
\label{eq:arrhenius}
,\end{equation}
where $\gamma$ is set equal to zero as the reaction channels are barrierless. The parameters were obtained from the best fit of the rate coefficients as a function of temperature, extrapolated from experimental data (grey and red curves in Fig. \ref{fig:ratesP+_C2H2}). For best fitting of the \ce{HCCP+} plus H channel the temperature values have been separated into three subsets.

\begin{table}[ht!]
\caption{Modified Arrhenius parameters for the rate coefficients of the reaction \ce{P+} plus \ce{C2H2}.}                 
\label{table:param}    
\centering                       
\begin{tabular}{l c c c}      
\hline\hline               
\noalign{\smallskip}
\textbf{Products} & \textbf{T range (K)} & $\alpha$($\times10^{-10}$) cm$^3$s$^{-1}$ & $\beta$ \\  
\noalign{\smallskip} 
\hline\hline 
 \noalign{\smallskip} 
   \ce{HCCP+} + \ce{H} & 10 - 1000 & 10.641 & -0.05596 \\
   \noalign{\smallskip}
                       & 1000 - 3000 & 10.909 & -0.07509 \\
                       & 3000 - 5000 & 11.711 & -0.10644 \\
 \noalign{\smallskip}
 \hline
 \noalign{\smallskip}
    \ce{CCP+} + \ce{H2} & 10 - 5000 & 0.2570 & -0.07299 \\
 \noalign{\smallskip}    
\hline\hline                                 
\end{tabular}
\tablefoot{The rate coefficients, $k(T)$, for the two open channels (column 1) are given in terms of $\alpha$ (column 3) and $\beta$ (column 4) parameters of the modified Arrhenius Eq. (\ref{eq:arrhenius}). The rates are valid for the range of temperatures indicated in column 2 and were obtained from best fits of $k(T)$ extrapolated from experimental data via Eq. \ref{KdeT}.
}
\end{table}

\section{Discussion}
\subsection{Rate coefficients}
The total rate determined from our fitted CSs differs slightly from a pure Langevin trend ($k=1.16\times10^{-9}$ cm$^3$s$^{-1}$ and independent of $T$, estimated using a polarizability of 3.487 \AA$^3$ for \ce{C2H2} \citep{Olney1997, NISTCCC} by showing an increase with decreasing temperature. At 300 K our total rate is in good agreement with the rate measured using SIFT , $1.3\times10^{-9}$ cm$^3$s$^{-1}$ at 300 K \citep{Smith_1989, Adams_1990}. 
It appears to however be slightly lower in our determination ($1.1\times10^{-9}$ cm$^3$s$^{-1}$ at 300 K), which may indicate that the population
of the fine levels of \ce{P+}(\ce{^3P}) is likely different in their experiment.
Indeed, in SIFT experiments performed at higher pressure, the fine levels are expected to be thermalized at 300 K, with most of the population in the \ce{^3P_0} level which will correlate to the most reactive triplet PES.
This may also be the reason why the \ce{CCP+} product was not observed in SIFT experiments.
In our study however, \ce{^3P_0}, \ce{^3P_1} and \ce{^3P_2} are expected to be equally populated, and consequently, reaction occurs on the three triplet PESs.
From Fig. \ref{fig:PECs}, only the PEC of the ground triplet state in the entrance channel is compatible with a high capture probability (direct access to a deep well without barrier). The other two triplet states can only form a charge-quadrupole Van der Waals complex where P$^+$ is located too far to be able to form a bond
and the curves suggest the presence of large barriers to access the region where bonds can form.
In principle, those two states are thus not expected to react at low collision energy, and in this case the CS at low energy would be close to $\sigma_{max}/3$ (red curve in Fig. \ref{fig:totCS}), and not 0.85$\sigma_{max}$, as we found in our experiment.
This implies that reactions occurring on the two higher triplet states also exhibit a rather high probability of reaction, and it is a strong indication that intersystem crossing (ISC) may play an important role.
There are two arguments in favour of this interpretation.
First, the formation of \ce{CCP+} is only exothermic in its singlet ground electronic state. The lowest isomer of \ce{CCP+} in a triplet state is endothermic by more than 0.5 eV (see Table \ref{tab:enthalpy_P_C2H2}). Since we observe it in our experiment at low collision energy, ISC has to occur.
Second, the two PECs of the $1^1A'$ and $2^1A''$ cross the PECs of  $1^3A'$ and $2^3A''$, and both crossings occur close to the bottom of the wells of \ce{P+}-H$_2$C$_2$ Van der Waals complex in each of the triplet states. 
The position of these crossings allows to increase drastically the ISC probabilities.
Indeed, being supported by a charge quadrupole interaction, the Van der Waals complexes are rather stable ($\approx 0.5 eV$). Even if they do not react directly, if formed, these complexes may thus survive a certain time before dissociating back to reagents, and the longer they survive, the larger becomes the probability to undergo an ISC.
Since the lifetime of the complexes increases with decreasing collision energy, we can expect that ISC will gain efficiency at lower collision energy.
This qualitative feature also explains why the measured CS is closer to 
$\sigma_{max}/3$ above 2 eV, and progressively increases while diminishing the collision energy until it reaches 0.85$\sigma_{max}$. 

In addition to the formation of \ce{CCP+}, another discrepancy between our results and those by \citet{Smith_1989, Adams_1990} using a SIFT apparatus concerns the adduct, \ce{H2CCP+}, formed as a minor channel (BR=5\%) in SIFT experiments. The formation of \ce{H2CCP+}, that is not  observed in our experiment, is likely due to the collisional stabilization enhanced by the intrinsically higher pressures of the SIFT experiment rather than to a radiative association. 
The formation of \ce{H2CCP+} by the title reaction is thus not expected to occur in the ISM.

\subsection{Charge transfer channel}

From Table \ref{tab:enthalpy_P_C2H2}, the charge transfer channel to produce C$_2$H$_2^+$ should open around 0.9 eV, but appears to be significant in our experiment only at energies $>2$ eV. Again, this can be rationalized simply by looking at the PECs of the entrance channel shown in Fig. \ref{fig:PECs}.
The charge transfer product channel, P$(^4S)$+\ce{C2H2+}($X^2\Pi)$, correlates to the $2^3A'$ and $3^3A''$ states, and lies slightly below the P$^+(^1D)$+C$_2$H$_2$($X^1\Sigma_g^+)$ entrance channel.
Clearly, an avoided crossing between the states $2^3A''$ (correlated to P$^+$) and $3^3A''$ (correlated to P) is observed around 2 \AA,  allowing population transfer between these two states. Similarly, the clear distortion in the repulsive part of the $^3A'$ potentials indicate couplings between those states.
Both $2^3A'$ and $3^3A''$ present a barrier to dissociation of about 2.8 eV and 2.5 eV, respectively, but these barrier heights are found considering a rigid structure of \ce{C2H2} in the {\it ab initio} calculations, and are expected to be lower if those states were optimized. 
The clear increase of the CS above 2 eV thus suggest that the charge transfer becomes efficient when the reaction takes place on these two states. On the contrary, the ground triplet state is not expected to offer an efficient mechanism for the charge transfer.
Finally, for collision energies above 7 eV, the charge transfer channel  becomes the principal reactive channel (see inset in Fig. \ref{fig:totCS}) and represent more than 80\% of products above 9 eV. This indicates that charge transfer also becomes efficient on the lowest triplet state.

\subsection{Additional isomers/electronic states for \ce{CCP+}}
\ce{P+}(\ce{^3P}) should have 3 associated electronic states which can reach the first three triplet states of \ce{CCP+}, if the reaction takes place adiabatically.
In each electronic states, linear isomers  of the type \ce{CCP+} and T-shape isomers of the type \ce{CPC+} are found stable, their values being provided in Table \ref{tab:enthalpy_P_C2H2}. 
Considering the calculated energetics of these different isomers, their contribution is expected to appear only at high collision energies, namely above 0.58 eV. Although a weak and broad feature can be tentatively identified between $\sim0.5$ and $\sim1.2$ eV in the experimental CS measured at $m/z$ 55 (blue circles in Fig. \ref{fig:P_C2H2_ecm_22.6eV}), the statistics on the data is insufficient to assign this feature to the appearance energy threshold of the \ce{CPC+} and \ce{CCP+} isomers in their lowest triplet state. 
A clearer change of tendency is observed above 3 eV, where the CS of formation of \ce{CCP+} start to increase with increasing collision energy, in good accord with the energy of the third triplet state of \ce{CPC+} being estimated to 3.33 eV. 
Of course, those are simply qualitative remarks, as other factors, such as non-adiabatic and spin-orbit couplings, barrier heights and dynamical properties of the reactions on the different PESs, are also expected to contribute to shape the CS.

As for the \ce{HCCP+} product, no special features are observed at the opening of its first excited state. 
However, the energy threshold of this state appears in a range in which its formation CS drops drastically in favour of the increase in the CSs for the charge transfer and the formation of \ce{CCP+}.

It is interesting to note that such less stable isomers seem to contribute more than the lower energy ones to the total CS at high collision energies. 
This can be reconciled by taking into account that the entrance channel for the \ce{P+} plus \ce{C2H2} reaction energetically favours a T-shape approach geometry, as also clearly shown by the calculations of \citet{cimas_computational_2012}, where \ce{P+} attacks the triple bond forming an initial cyclic intermediate (denominated I$_1$) which lies $\sim61$ kcal mol$^{-1}$ (2.65 eV) below the reactants (on the triplet PES) and at even lower energy, $\sim140$ kcal mol$^{-1}$ (6.0 eV), on the singlet PES correlating with \ce{P+}(\ce{^{1}D}).
From such adducts, to reach the linear geometry of the \ce{CCP+}($^1\Sigma$) plus \ce{H2} product, it is necessary to follow a complex pathway where H and P atoms would have to move on the PES. This is confirmed by the pathway on the singlet PES shown in Fig. 4 of \citet{cimas_computational_2012} where a first H shift is required to go from I$_1$ to I$_2$ (\ce{P-C-CH2+} geometry) with a barrier of 93 kcal mol$^{-1}$ (4.0 eV) and then the ejection of an \ce{H2} is also hampered by a second barrier of similar value.   
On the other hand, reaching the T-shape geometry of the \ce{CPC+} (plus \ce{H2}) higher energy products requires only H atoms to move on the PES, and this is an "easier" pathway. 

\subsection{Astrochemical implications}

\begin{figure}[h]
   \centering
   \includegraphics[width=\hsize]{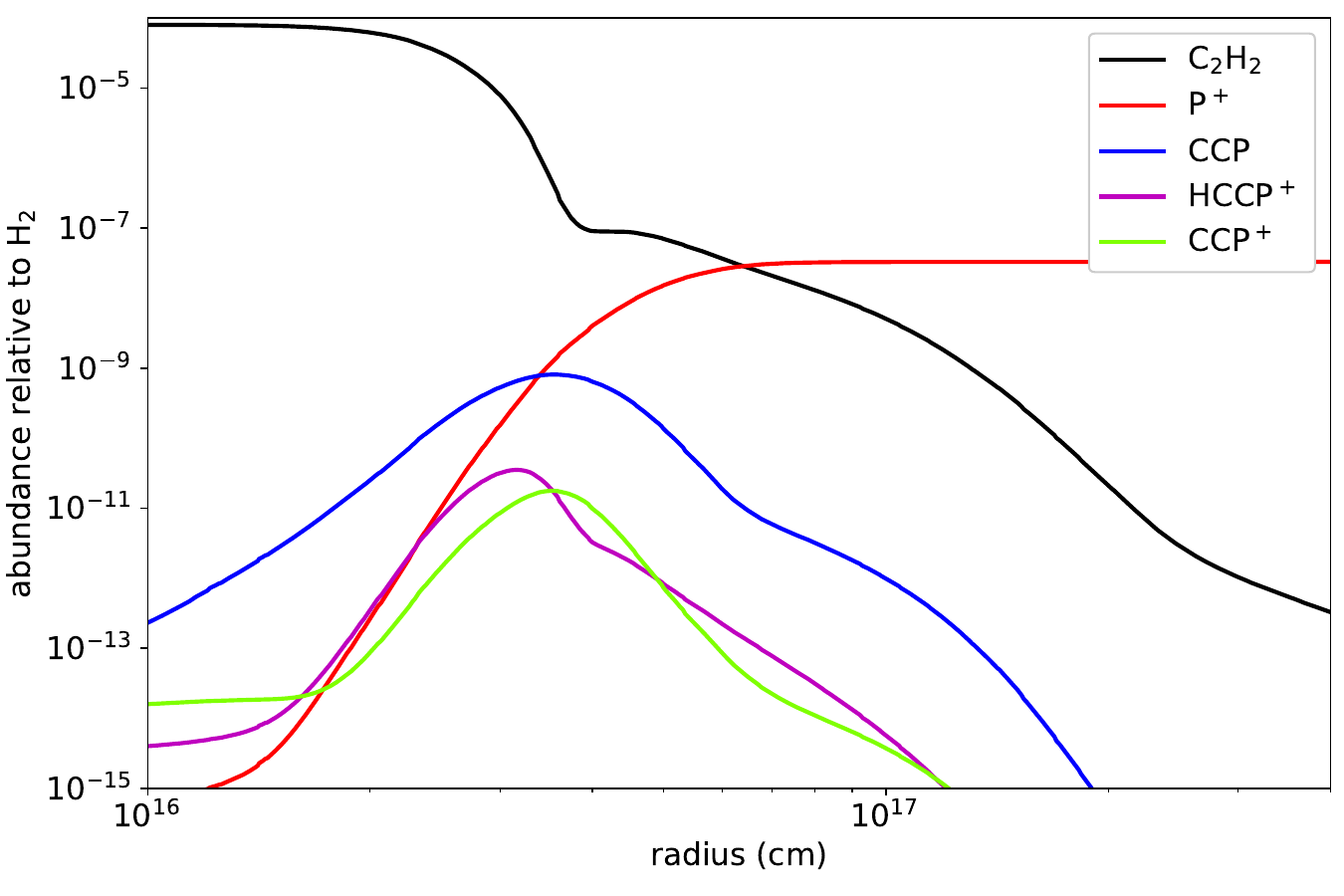}
   \caption{Modelled relative abundances of phosphorus species related to the title reaction in the circumstellar envelope IRC\,+10216 as a function of the distance from the star.}
         \label{fig:model}
   \end{figure}

In order to see the impact of the studied reaction on astronomical environments where phosphorus chemistry is known to be important, we carried out chemical modelling calculations of the carbon-rich circumstellar envelope IRC\,+10216, where many P-bearing molecules have been detected. The core of the model is based on \cite{Agundez2017}, where we included the reaction studied in this work with the rate coefficient expression given in Table~\ref{table:param}.
 The calculated abundances of several key phosphorus species is shown in Fig.~\ref{fig:model} as a function of radial distance to the star.
 According to the model, the reactants \ce{P+} and \ce{C2H2} are not co-spatial since the former is mostly present in the outer parts and the later in the inner ones. However, there is a region located around 3\,$\times$\,10$^{16}$ cm where both species are present with non-negligible abundances, leading to the formation of the main product, \ce{HCCP+}. Although the chemistry in the outer envelope of IRC\,+10216 is mainly neutral, the title reaction is the main doorway to P-bearing cations in this source as it involves two rather abundant reactants. It is the main source of \ce{HCCP+} in IRC\,+10216. The secondary product, \ce{CCP+}, is predicted to be mostly formed by the ionization of the neutral radical \ce{CCP} rather than by the title reaction, but is associated with large uncertainties since most of the chemistry related to neutral \ce{CCP} is not known, and rough estimates of the rates are employed in the model.
 Nevertheless, from the actual predictions, it is remarkable that both cations, \ce{HCCP+} and \ce{CCP+}, are predicted with a similar peak abundance of (1-4)\,$\times$\,10$^{-11}$ relative to \ce{H2}. A good test of the model would be to detect these two cations and derive their abundances. However, to the best of our knowledge, their rotational spectra have not been measured in the laboratory. The related species \ce{CCP} has been detected in IRC\,+10216 with an abundance of $\sim$10$^{-9}$ \citep{Halfen2008}, in agreement with the prediction of the model.

In other astronomical environments such as interstellar clouds, the title reaction could play an important role as it involves two potentially abundant species. \ce{P+} is expected to be the main form of phosphorus in diffuse clouds and the starting point to the chemistry of phosphorus once the cloud evolves to a dense and UV-shielded phase, and acetylene is probably an abundant hydrocarbon although its lack of dipole moment prevents to directly detect it. In these environments, the dissociative recombination of HCCP$^+$ with electrons can be an important source of neutral species, such as CCP and CP \citep{Millar_1991,Chantzos2020}.

\section{Conclusions}

This study reports an experimental and computational investigation of the ion-molecule reaction between \ce{P+} and acetylene. Absolute CSs and BRs were measured in the collision energy range 0.1 - 10 eV in the centre-of-mass frame, using a guided ion beam set-up coupled to VUV photoionization via synchrotron radiation, to generate \ce{P+} ions in their ground electronic state, $^{3}P$.
Electronic structure calculations were carried out to determine reaction enthalpies across multiple triplet and singlet electronic states, and a refined capture model, improved from the standard Langevin one, is proposed to properly account for the formation of the intermediate complex and to describe the energy dependence of the reactive CS at ultra low collision energies. From the experimental CSs, and with the insights provided by the theoretical treatment, rate coefficients as a function of temperature, $k(T)$, were derived in the 10-5000 K range. 
The main results can be summarized as follows
\begin{itemize}
\item 
The reaction \ce{P+} plus \ce{C2H2}, at the lowest collision energies explored by our experiment, leads mostly to \ce{HCCP+} plus H (with a BR = 99 \%), while the \ce{CCP+} plus \ce{H2} channel accounts for 1\%, a value that is a factor 10 larger than what predicted by a previous theoretical study \citep{cimas_computational_2012} 
\item 
The obtained total reaction rate coefficient at 300 K ($1.1\times10^{-9}$ cm$^3$s$^{-1}$) is in good agreement with SIFT measurement ($1.3\times10^{-9}$ cm$^3$s$^{-1}$), but differ from a pure Langevin trend by showing an increase with decreasing temperature: $1.3\times10^{-9}$ and $8.9\times10^{-10}$ cm$^3$s$^{-1}$ at 10 and 5000 K, respectively.
\item 
Chemical modelling calculations of the carbon-rich circumstellar envelope IRC\,+10216 show that, although the chemistry in the outer envelope is mainly neutral, the title reaction is the main doorway to P-bearing cations and the principal pathway to \ce{HCCP+} in this source.
\item
The relevant astrochemical isomer produced from the title reaction is \ce{HCCP+} in its linear structure, since the cyclic product is endothermic.
\item 
The reaction of \ce{P+} with \ce{C2H2} is the main pathway to ion phosphorus chemistry in circumstellar envelopes around carbon stars, leading to the formation of \ce{HCCP+}. In interstellar clouds, the dissociative recombination of its main product, \ce{HCCP+}, with electrons can be an important source of neutral species such as CP and CCP.
\end{itemize}   

\begin{acknowledgements}
The authors acknowledge SOLEIL for providing synchrotron radiation necessary for the experiments and Laurent Nahon and his team for assistance in using the DESIRS beamline under proposal 20231311.
The research leading to these results has received funding from MCIU (Spain) and FEDER (UE) under grant PID2021-122549NB-C21, PID2021-122549NB-C22  and PID2024-155352NB-C22. The computing time on HPC DRAGO (CSIC) is also acknowledged.
DA acknowledges financial support from MUR PRIN 2020 project n. 2020AFB3FX "Astrochemistry beyond the second period elements". MM acknowledges the Department of Physics at the University of Trento for co-financing a PhD scholarship, and the COST Action CA22133 - PLANETS for the financial support while carrying out the experiments at the SOLEIL synchrotron. MP acknowledges the COST Action CA21126 - Carbon molecular nanostructures in space (NanoSpace) for financial support while conducting the experiments at synchrotron SOLEIL. The  COST Action CA21101 (Cozy) is also acknowledged for its support and facilitation of collaborations.

\end{acknowledgements}

\bibliographystyle{aa}
\bibliography{RefsPC2H2}

\begin{appendix}
\nolinenumbers

\section{Quantification of the absence of metastable excited states in the \ce{P+} Ion Beam}
\label{sec:appendix_A}

The reaction of \ce{P+} with \ce{N2} has been used to verify the presence of the metastable excited states \ce{^{1}D} and \ce{^{1}S} in the ion beam generated by dissociative photoionization of \ce{PCl3}, thus allowing to chose the optimal photon energy for a state-selection of the ground state of the parent ion. 
The reaction enthalpies for the different electronic states of \ce{P+} with \ce{N2} according to reaction (\ref{eq:PN+}) are shown in Table \ref{tab:P+_N2}:

\begin{equation}
\label{eq:PN+}
\begin{aligned}
\ce{P+ + N2} &\rightarrow & \ce{PN+}(X^2\Sigma^+) && (m/z & & 45) &&  + & & \ce{N} (^4S)
\end{aligned}
\end{equation}

\begin{figure}[h]
   \centering
   \includegraphics[width=1.0\hsize]{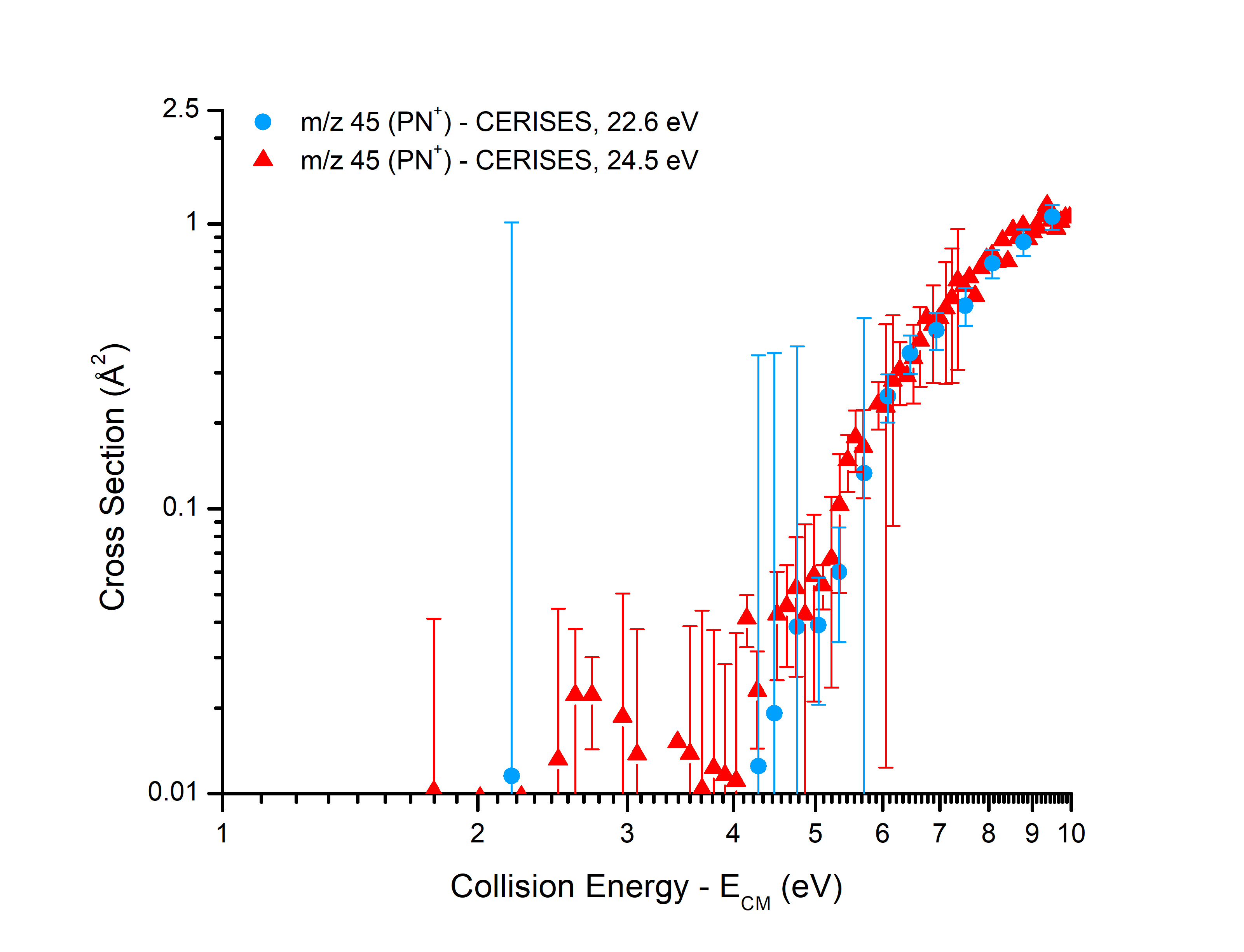}
      \caption{Experimental cross sections for the reaction of \ce{P+} with \ce{N2} as a function of the collision energy, $E_{CM}$. The blue and red circles are the profile obtained at $E_\mathrm{phot}$ = 22.6 eV and 24.5 eV, respectively.} 
      \label{fig:PN+_ecm}
   \end{figure}

Since reaction (\ref{eq:PN+}) from \ce{P+}(\ce{^{3}P}) is endothermic by over 4 eV, it is expected to give no products except at high collision energies, and it is a viable method to assess excited state contamination, since in the presence of \ce{^{1}D} and/or \ce{^{1}S} states, lower energy thresholds would be expected for the \ce{PN+} product.
Absolute CSs as a function of collision energy for reaction (\ref{eq:PN+}) measured at two different photon energies $E_\mathrm{phot}$ = 22.6 and 24.5 eV are compared in Fig.~\ref{fig:PN+_ecm}.

\begin{table}[h]
\centering
\caption{
Enthalpies of reaction ($\mathbf{\Delta}H^0$) for \ce{P+} (in ground and lowest-energy metastable excited states) with \ce{N2}, derived from literature data.}
\label{tab:P+_N2}
\begin{threeparttable}
\begin{tabular}{l c}
\hline\hline
\noalign{\smallskip}
\textbf{Reagent ion} & \textbf{$\Delta_{rxn} H^0$ (eV, 298 K)\tnote{a}} \\
\noalign{\smallskip}
\hline\hline
\noalign{\smallskip}
\ce{P+ (\ce{^{3}P})} & $+4.10\pm0.05$ \\
\noalign{\smallskip}
\hline
\noalign{\smallskip}
\ce{P+ (\ce{^{1}D})} & $+2.99\pm0.06 $ \\
\noalign{\smallskip}
\hline
\noalign{\smallskip}
\ce{P+ (\ce{^{1}S})} & $ +1.43\pm0.04 $ \\
\noalign{\smallskip}
\hline
\noalign{\smallskip}
\hline
\end{tabular}

\begin{tablenotes}
\footnotesize
\item[a] The values were estimated from data available at \citet{ATcT, NISTchem}.
\end{tablenotes}
\end{threeparttable}
\end{table}

Both data exhibit distinct energy thresholds, above which the reaction occurs, with CSs increasing as a function of collision energy. For data at $E_\mathrm{phot}$ = 22.6 eV the estimated energy threshold is $4.0\pm0.1$ eV, while for data $E_\mathrm{phot}$ = 24.5 eV a lower value of $3.5\pm0.3$ eV is observed.
According to the reaction enthalpies reported in Table \ref{tab:P+_N2}, the threshold for reaction of \ce{P+}(\ce{^{1}D}) is expected at $\sim3.0$ eV, while that for the reaction of \ce{P+}(\ce{^{1}S}) is at $\sim1.4$ eV. Since no reaction is observed with an AE around 1.4 eV, the presence of the metastable \ce{^{1}S} state in the beam can be excluded, while the lowering of the energy threshold at $E_\mathrm{phot}$ = 24.5 eV can be explained with the production of a certain, although small, amount of \ce{P+}(\ce{^{1}D}) at this photon energy.
In fact, based on the analysis of dissociative photoionization of \ce{PCl3} reported in \citep{de_la_fuente_michielan_2026} and on the calculated dissociation thresholds for production of the metastable states via the ion-pair dissociation channel and the "complete" dissociation channels (see Sec. \ref{sec:exp_met}) when tuning the photon energy at 22.6 eV we are above the thermodynamic threshold for  production of \ce{^{1}D} via both dissociation channels (AEs at 16.6 eV and 21.6 eV), while for the \ce{^{1}S} state, we are above the ion-pair dissociation channel (AE 18.2 eV) but below the "complete" dissociation channel (AE 23.2 eV). Hence, we can conclude that the ion pair dissociation channel does not produce the metastable states \ce{^{1}D} and \ce{^{1}S} (or at least not in sufficient quantity to observe their reactivity). 
Experiments performed at the higher photon energy $E_\mathrm{phot}$ = 24.5 eV indicate that, even though the "complete" dissociation channel producing \ce{P+}(\ce{^{1}S}) is open, the absence of reactivity may indicate either that \ce{^{1}S} is not produced, or that it rapidly relaxes to the \ce{^{1}D} state. In conclusion, dissociative photoionization of \ce{PCl3} at $E_\mathrm{phot}$ = 22.6 eV produces exclusively \ce{P+} in its ground electronic state.

\end{appendix}

\end{document}